\documentclass[a4paper,fleqn,usenatbib]{mnras}
\usepackage[T1]{fontenc}
\usepackage{ae,aecompl}
\usepackage{graphicx}	
\usepackage{amsmath}	
\usepackage{amssymb}	

\usepackage[utf8]{inputenc}
\usepackage{natbib}
\usepackage{epsfig}
\usepackage{appendix}
\usepackage{rotating}

\title[]
{UV photoprocessing of NH$_3$ ice: photon-induced desorption mechanisms}

\author[R. Mart\'in-Dom\'enech et al.]
{R. Mart\'in-Dom\'enech,$^{1}$ 
G. A. Cruz-D\'iaz,$^{2,3}$ 
G. M. Mu\~noz Caro,$^{4}$ 
\\
$^{1}$Harvard-Smithsonian Center for Astrophysics, Cambridge, MA 02138, USA\\
$^{2}$Bay Area Environmental Research Institute, Petaluma, CA 94952, USA\\
$^{3}$NASA Ames Research Center, Moffet Field, Mountain View, CA 94035, USA\\
$^{4}$Centro de Astrobiolog\'ia (INTA-CSIC). Ctra de Ajalvir, Km. 4, Torrej\'on de Ardoz, 28850 Madrid, Spain\\
}

\date{Accepted XXX. Received YYY; in original form ZZZ}

\pubyear{2017}

\begin{document}
\label{firstpage}
\pagerange{\pageref{firstpage}--\pageref{lastpage}}
\maketitle

\begin{abstract}
Ice mantles detected on the surface of dust grains toward the coldest regions of the interstellar medium can be photoprocessed by 
the secondary ultraviolet (UV) field present in dense cloud interiors. 
In this work, we present UV-irradiation experiments under astrophysically relevant conditions of pure NH$_3$ ice samples in an 
ultra-high vacuum chamber where solid samples were deposited onto a substrate at 8 K. 
The ice analogs were subsequently photoprocessed with a microwave-discharged hydrogen-flow lamp. 
The induced radiation and photochemistry led to the production of H$_2$, N$_2$ and N$_2$H$_4$. 
In addition, photodesorption to the gas phase of the original ice component, NH$_3$, and two of the three detected photoproducts, 
H$_2$ and N$_2$, was observed thanks to a quadrupole mass spectrometer (QMS). 
Calibration of the QMS allowed quantification of the photodesorption yields, leading to $Y_{pd}$ (NH$_3$) 
= 2.1$^{+2.1}_{-1.0}$ x 10$^{-3}$ $\frac{\text{molecules}}{\text{incident photon}}$, which remained constant during the whole experiments,  
while photodesorption of H$_2$ and N$_2$ increased with fluence, pointing toward an indirect photodesorption mechanism involving 
energy transfer for these species. 
Photodesorption yield of N$_2$ molecules after a fluence equivalent to that experienced by ice mantles in space 
was similar to that of the NH$_3$ molecules ($Y_{pd}$ (N$_2$) = 1.7$^{+1.7}_{-0.9}$ x 10$^{-3}$ $\frac{\text{molecules}}{\text{incident photon}}$). 
\end{abstract}




\section{Introduction}
\label{intro}

Infrared observations, mainly performed by space telescopes like the Infrared Space Observatory (ISO), Spitzer, and AKARI, 
have revealed the presence of ice mantles on the surface of interstellar dust grains in dense interstellar clouds 
(see, e.g., \cite{gibb04,oberg11}, and references therein). 
Water is the major component of the ice mantles, but five more species are confirmed to be present with varying 
abundances relative to H$_2$O: CO, CO$_2$, CH$_3$OH, NH$_3$, and CH$_4$; and several more have been tentatively detected 
(\cite{boogert15} and references therein). 

Ice molecules are processed by cosmic rays and the secondary ultraviolet (UV) field produced in the interior of dense clouds  
when the cosmic rays interact with the gas-phase H$_2$ molecules (\cite{cecci92,shen04}).  
In particular, UV photons induce photochemical reactions by photolysis of absorbing molecules and recombination of the  
produced radicals (see \citet{oberg16} and references therein),  
and also photodesorption processes, which represent one of the proposed non-thermal desorption mechanisms necessary to explain the 
presence 
of gas-phase molecules in the coldest regions of the interstellar medium (ISM), where they should be completely depleted 
forming these ice mantles (e.g., \citet{willacy00,bergin01}). 

\smallskip
Depending on whether the photodesorbing molecule is the same ice molecule that absorbs the UV photon 
(or at least  a fragment of the molecule), or a different molecule, one can distinguish between direct photodesorption mechanisms, 
and photodesorption mechanisms involving energy transfer, respectively. 

In the case that the photon is absorbed by a surface ice molecule that does not subsequently photodissociate, 
the absorbing molecule can directly photodesorb to the gas phase. 
This mechanism has been proposed by \cite{vanhemert15} for the photodesorption of CO molecules from a pure CO ice 
based on molecular dynamics simulations. However, in previous works this process had been proven to be inefficient during experimental simulations 
with CO and N$_2$ ices when compared to indirect photodesorption mechanisms 
(see, e.g., \cite{guillermo10,bertin12,bertin13}), although it may have a contribution in other cases (\cite{dupuy17}). 
Alternatively, if the absorbing surface molecule photodissociates, 
the resulting photofragment can desorb provided that it is formed with enough kinetic energy, 
or it may recombine with another fragment, leading to the formation of a photoproduct that could also photodesorb  
thanks to the excess energy of the parent photofragments and/or the exothermicity released during the recombination reaction 
(see, e.g., \cite{andersson08,fayolle13,fillion14,bertin16}). 
Since the direct photodesorption of photoproducts through this mechanism 
(which is referred to as photochemical desorption or photochemidesorption in previous works; \cite{martin16,gustavo16}) 
takes place immediately after their formation, photoproducts cannot be accumulated on the ice surface prior to their photodesorption, 
and the measured photodesorption yield during experimental simulations remains constant with fluence. 

On the other hand, if the photon is absorbed by a molecule below the ice surface leading to its electronic excitation, 
the energy can be redistributed or transferred to a surface molecule
\footnote{The transfer of energy could also take place between molecules on the surface of the ice.} 
that could break the intermolecular bonds and photodesorb. 
This process is usually known as desorption induced by electronic transitions (DIET) 
followed by subsequent energetic transfer
\footnote{The term DIET could, in principle, encompass all photodesorption processes, and does not necessarily imply energetic 
transfer. However these two concepts tend to be related in the 
literature (see \cite{fayolle11,bertin12}). 
In addition, when the absorbing and the photodesorbing 
molecules belong to different species during energy-transfer mediated processes, this mechanism is sometimes referred to as indirect DIET or photon-induced co-desorption, 
although no distinctions are made, in general.}  
(\cite{fayolle11,fayolle13,bertin12,bertin13,fillion14}), and it is referred to as kick-out photodesorption in \cite{vanhemert15}. 
In the case that the absorbing molecule dissociates, the excess energy can be transferred by the resulting photofragment 
to a surface molecule, leading to its photodesorption. 
When the photofragment is an H atom, it can diffuse through the ice and transfer its momentum to a surface molecule. 
This is known as kick-out photodesorption\footnote{Not to be confused with the kick-out photodesorption in \cite{vanhemert15}.} 
(\cite{andersson08}). 
Alternatively, the excess energy from the recombination of the photofragments can also be transferred to a surface molecule leading 
to its photodesorption (\cite{andersson08,fillion14}). 
When a previously formed photoproduct photodesorbs through any of these indirect mechanisms, accumulation in the ice prior to their 
photodesorption leads to an increasing photodesorption yield with fluence, as measured during experimental simulations 
(\cite{martin15,martin16,gustavo16}). 
\smallskip

Experimental simulations aiming to study the photodesorption of molecules taking place during photoprocessing of pure ices made 
by the six species confirmed to be present in interstellar ice mantles are a necessary first step prior to the study of the 
photodesorption in more realistic multicomponent ice analogs. 
These studies have been reported for pure 
H$_2$O (\cite{gustavo17}, and references therein), 
CO (\cite{oberg07,oberg09b,guillermo10,fayolle11,bertin12,bertin13,chen14,guillermo16}),
CO$_2$ (\cite{oberg09b,bahr12,yuan13,fillion14,martin15}),
CH$_3$OH (\cite{oberg09a,bertin16,gustavo16}),
CH$_4$ (\cite{dupuy17}), 
and NH$_3$ (\cite{nishi84,loeffler10}) ices. 
In this work, we present a series of experiments simulating the UV photoprocessing of pure NH$_3$ ices focused on the subsequent 
photodesorption of ice molecules. 
Irradiation was carried out with a multiwavelength UV lamp, instead of a laser at a given wavelength used in \cite{nishi84,loeffler10}. 
In addition, we include the quantification of the different photodesorbing molecules measured directly from the gas phase, 
which was not reported in the previous works mentioned above,
thanks to the calibration of the quadrupole mass spectrometer (QMS) used in our experimental setup (presented in 
Section \ref{methods}) to detect the photodesorbing molecules. These results are presented in Section \ref{results}, and their 
astrophysical implications are discussed in Section \ref{imp}. Finally, the conclusions are summarized in Section \ref{conclusions}.

\begin{table*}
\tabcolsep 8pt
\begin{center}
\caption{Experimental parameters of the UV photoprocessing of pure NH$_3$ ices}
\label{exps}
\begin{tabular}{c c c c c c c}
\hline
\hline
Experiment & N$_{initial}$(NH$_{3}$) & Fluence$_{incident}$ & Fluence$_{absorbed}^{a}$ & Dose$_{incident}^{b}$ & Dose$_{absorbed}^{a}$ & Heating rate\\
& ($\times$10$^{15}$ molecules cm$^{-2}$) & \multicolumn{2}{c}{($\times$10$^{18}$ photons cm$^{-2}$)} & 
\multicolumn{2}{c}{(photons molecule$^{-1}$)} & (K/min)\\
\hline
1 & 237 & 2.16 & 1.32 & 9.11 & 5.58 & 2\\
2 & 835 & 2.88 & 2.78 & 3.45 & 3.33 & 2 \\
3 & 912 & 2.88 & 2.80 & 3.16 & 3.07 & 2\\
4 & 913 & 0.72 & 0.70 & 0.79 & 0.77 & - \\
5 & 1882 & 1.24 & 1.24 & 0.66 & 0.66 & - \\
6 & 2264 & 13.27 & 13.27 & 5.86 & 5.86 & 2\\
\hline
\end{tabular}

\begin{list}{}
 \item $^{a}$ The number of absorbed photons has been calculated taking into account the initial composition of the ice, i.e., 
 the average UV absorption cross section for a pure NH$_{3}$ ice, provided by \cite{gustavo14a}.\\ 
 \item $^{b}$ Relative to the initial number of NH$_{3}$ molecules.\\ 
\end{list}
\end{center}
\end{table*}

\section{Experimental setup}
\label{methods}
The photodesorption of molecules from a pure NH$_3$ ice has been studied through a series of experimental simulations carried out 
with the InterStellar Astrochemistry Chamber (ISAC), an ultra-high-vacuum (UHV) setup located at the Centro de Astrobiolog\'ia,  
with a working pressure of $\sim$4 x 10$^{-11}$ mbar, similar to that found in the interiors of dense interstellar clouds.  
A brief description of the experimental setup and the protocol followed during the experimental simulations is provided below 
(see \cite{guillermo10,martin15,gustavo16,martin16} for more details). 
Pure amorphous ammonia ice samples were deposited from the gas phase onto a KBr window at 8 K used as the substrate, 
upon introduction of NH$_3$ (gas, Praxair 99.999\%) into the chamber. 
%
The deposited ices were subsequently irradiated using an F-type microwave-discharged hydrogen flow lamp (MDHL)   
with a vaccum-ultraviolet (VUV) flux of 2 $\times$ 10$^{14}$ photons cm$^{-2}$ s$^{-1}$ at the sample position 
(\cite{guillermo10}). 
The total irradiation time varied between 60 and 1100 min, leading to the incident fluences indicated in Table \ref{exps}. 
For a given fluence, the ice thickness determined the percentage of absorbed photons according to the VUV-absorption cross section  
reported in \cite{gustavo14a}\footnote{The average absorption cross section of pure NH$_3$ ice is 6.1 $\times$ 10$^{-18}$ cm$^{2}$ (\cite{gustavo14a}), 
which led to an absorption close to 100\% of incident photons in most of the experiments (see Table \ref{exps}).}, 
as well as the incident and absorbed dose of photons per molecule.  
The results in Section \ref{results} were not found to strongly depend on these parameters. 
The emission spectrum of the MDHL has been characterized in situ using a VUV spectrophotometer, and is reported in \cite{chen10,chen14, gustavo14a}.  
It is similar to the secondary UV field of 
dense cloud interiors calculated by \citet{gredel89}.  
%
Finally, the ice samples in experiments 1-3 and 6 were warmed-up following irradiation to room temperature at a rate of 2 K/min.  

During the experimental simulations, in situ Fourier-transform infrared (FTIR) transmittance spectroscopy was used to monitor the 
solid sample. 
IR spectra of the ices were collected after deposition, after every irradiation period, 
and every five minutes during warm-up, with a spectral resolution of 2 cm$^{-1}$. 
%
The initial column density $N$ in molecules cm$^{-2}$ of the NH$_3$ ices was calculated from the the optical depth ($\tau_{\nu}$) of the 
N-H stretching absorption IR band at $\sim$3300 cm$^{-1}$, using the formula:

\begin{equation}
N=\frac{1}{A}\int_{band}{\tau_{\nu} \ d\nu},
\end{equation}

\noindent where $A$ is the band strength in cm molecule$^{-1}$ 
(2.2 $\times$ 10$^{-17}$ cm molecules$^{-1}$; \cite{schutte96}). 
The ice thickness of the samples is usually expressed in monolayers (ML). One ML is assumed to be 10$^{15}$ molecules cm$^{-2}$. 

At the same time, all the desorbing species, including ammonia and the resulting products of the induced photochemical reactions, 
were detected in the gas phase by a Pfeiffer Prisma quadrupole mass spectrometer (QMS).  
The molecules reaching the QMS were ionized by $\sim$ 70 eV electron bombardment, which led to fragmentation following a given 
pattern. The main mass fragments coincided in this case with the molecular ions, and were used to monitor the presence of 
NH$_3$ (m/z = 17), and the photoproducts N$_2$ (m/z = 28), H$_2$ (m/z = 2), and N$_2$H$_4$ (m/z = 32) in the gas phase. 
The conversion from the integrated ion currents measured by the QMS for every irradiation period ($A (m/z)$) into photodesorbing 
column densities ($N (mol)$) was carried out using the equation: 

\begin{small}
\begin{equation}
N (mol) = \frac{A (m/z)}{k_{CO}} \cdot \frac{\sigma^{+} (CO)}{\sigma^{+} (mol)} \cdot \frac{I_{F} (CO^{+})}{I_{F} (z)} 
\cdot \frac{F_{F} (28)}{F_{F} (m)} \cdot \frac{S (28)}{S (m/z)} 
\label{eqmscal}
.\end{equation}
\end{small}

The meaning of the different parameters in Eq. \ref{eqmscal} is described in \cite{martin15,gustavo16,martin16,gustavo17}. 
The values used in this work are summarized in Table \ref{param}, and in the text. 
The constant $k_{CO}$ and the sensitivity of the QMS are regularly calibrated (see \cite{martin15} for more information on the 
calibration process). 


%
Since the pumping speeds in the ISAC setup are not the same for all molecules, $N (mol)$ from Eq.\ref{eqmscal} must be corrected 
to take into account the different pumping speed of the species of interest with respect to the CO molecules used as reference 
to extract $k_{CO}$:

\begin{equation}
N^{real} (mol) = N (mol) \cdot S_{rel} (mol) 
\label{eqmscorr}
,\end{equation}

\noindent where the relative pumping speed ($S_{rel} (mol)$) 
can be calculated as 

\begin{equation}
S_{rel} (mol) = 1.258 - 9.2 \cdot 10^{-3} \cdot M (mol)
\label{eqmspump}
,\end{equation}

\noindent 
(see \cite{kaiser95,martin16} for more information). 
Photodesorption yields are finally calculated dividing $N^{real} (mol)$ by the fluence.

\begin{table*}

\begin{center}
\caption{Values of the parameters in Eq. \ref{eqmscal} used to convert integrated QMS signals into photodesorbing column densities.  
Photodesorbing H$_2$ was not quantified due to contamination issues (see text).}
\label{param}
\begin{tabular}{cccc}
\hline
\hline
Factor&NH$_{3}$&N$_2$&CO\\
\hline
$\sigma^{+} (mol)$ (angstroms$^{2}$)$^{a}$&3.036&2.508&2.516\\
fragment&NH$_3^{+}$&N$_2^{+}$&CO$^{+}$\\
$m/z$&17&28&28\\
$I_{F} (z)$&1$^{b}$&1$^{b}$&1$^{b}$\\
&&&\\
&0.460$^{c}$&&\\
$F_{F} (m)$&0.514$^{d}$&0.933$^{a}$&0.949$^{a}$\\
&0.498$^{e}$&&\\
&&&\\
$k^{*}_{QMS} \cdot S (m/z)$\footnote{the ratio  $S (m/z)$/$S (28)$ is the same as the ratio $k^{*}_{QMS} \cdot S (m/z)$/$k^{*}_{QMS} \cdot S (28)$.} (A mbar$^{-1}$ $\AA^{-2}$)&1.78 x 10$^{15}$&1.03 x 10$^{15}$&1.03 x 10$^{15}$\\ 
$S_{rel} (mol)^{f}$&1.1&1.0&1.0\\
\hline
\end{tabular}

\begin{list}{}
\item $^{a}$ Extracted from the online database of the National Institute of Standard and Technologies (NIST).\\
\item $^{b}$ We assumed that no double ionization of the molecules took place in the QMS.\\
\item $^{c}$ Calculated during the deposition in experiment 1\\
\item $^{d}$ Calculated during the deposition in experiments 2,3,4, and 6\\
\item $^{e}$ Calculated during the deposition in experiment 5\\
\item $^{f}$ Calculated with equation \ref{eqmscorr}.\\
\end{list}
\end{center}
\end{table*}

\section{Experimental results and discussion}
\label{results}

\subsection{Photon-induced chemistry of pure NH$_{3}$ ice}

\begin{figure*}
\centering
\includegraphics[scale=0.45]{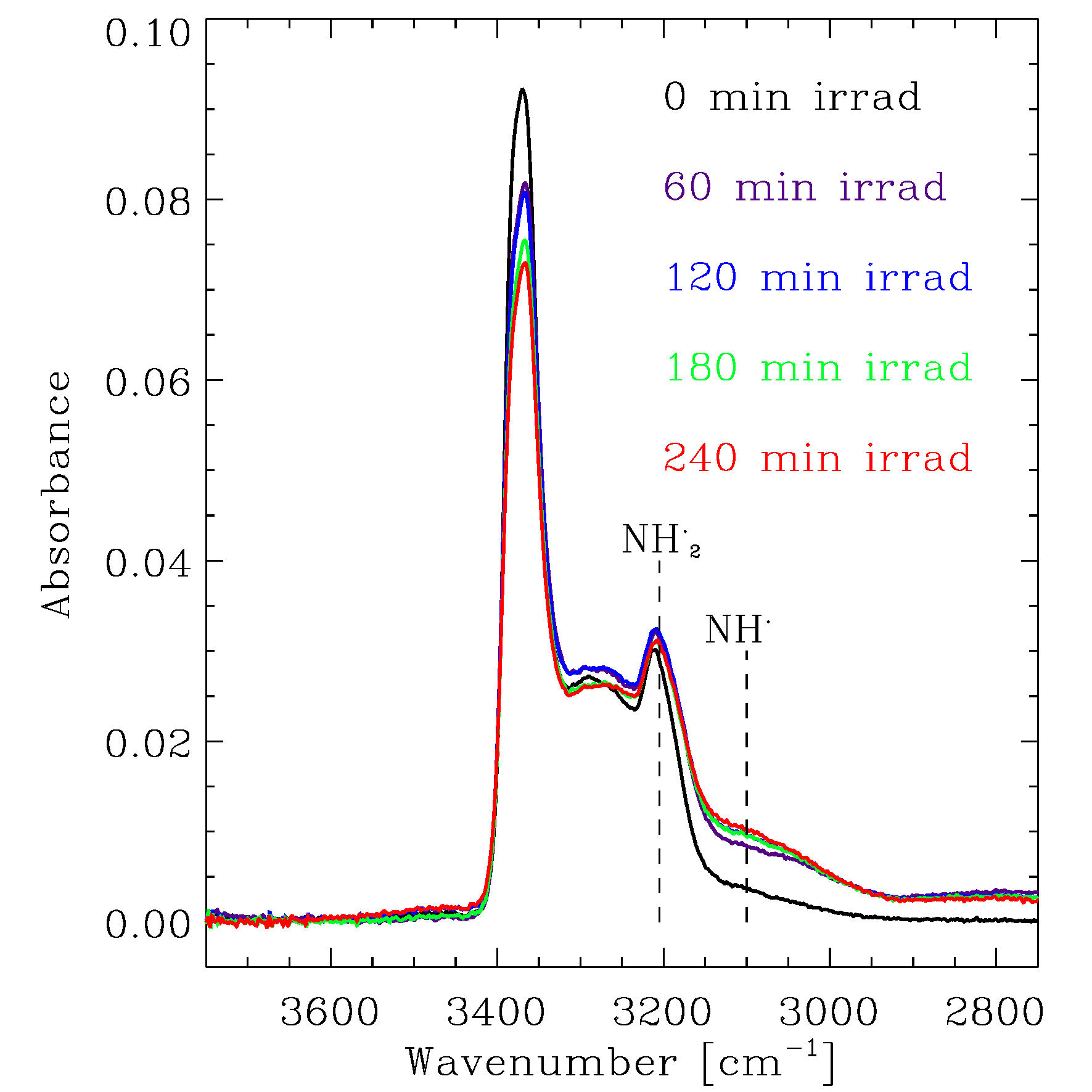}
\includegraphics[scale=0.45]{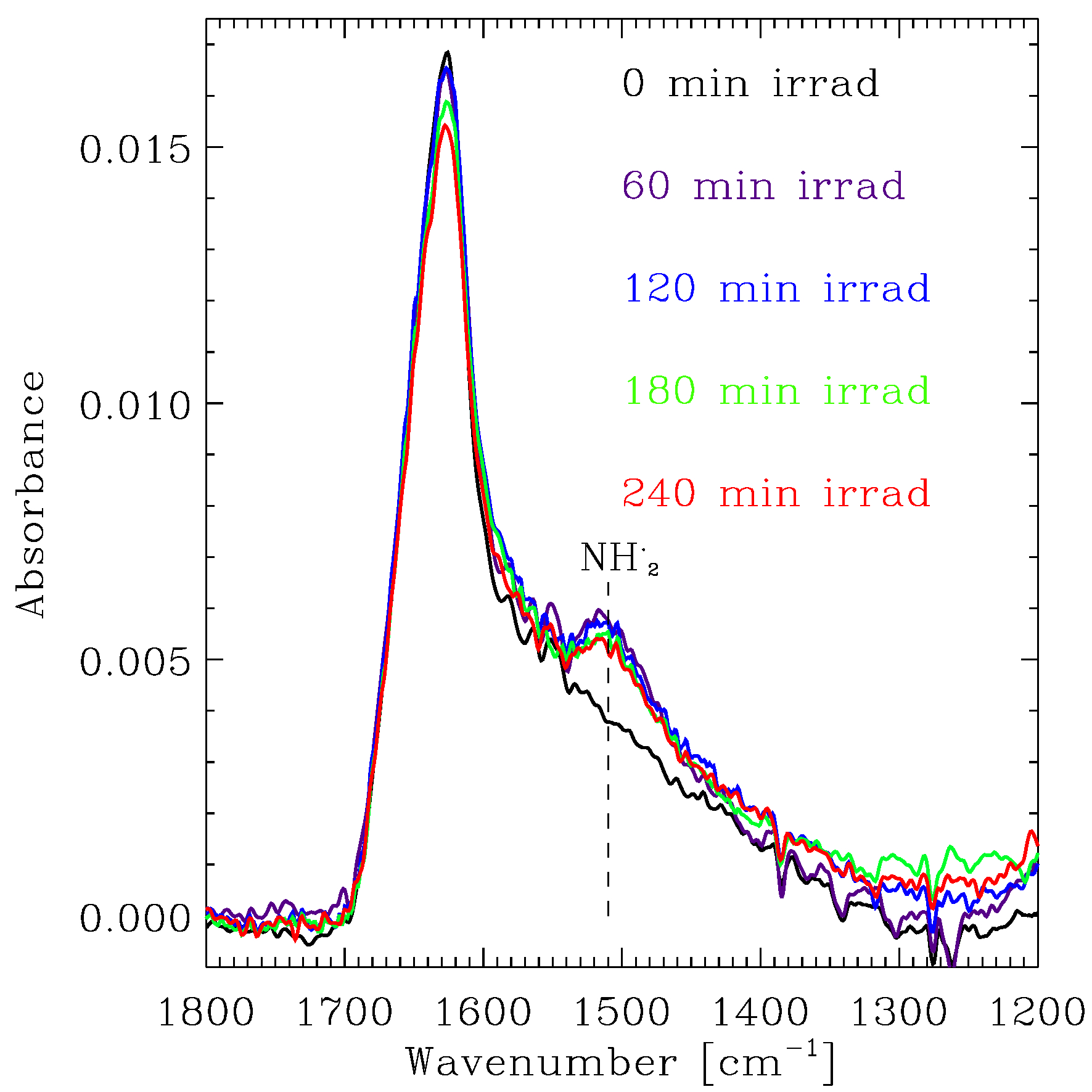}
\caption{IR spectra of the ice sample after every irradiation interval during photoprocessing in experiment 3 (see Table \ref{exps}). 
Similar results were found for the rest of the experiments. 
\textit{Left:} IR band corresponding to the N-H stretching modes of the NH$_{3}$ 
molecules. The decrease in the IR absorbance is due to the photodissociation and photodesorption of NH$_{3}$ molecules during 
irradiation. The weak shoulder above 3000 cm$^{-1}$ corresponds to the N-H stretching mode of the NH$^{\cdot\cdot}$ radical. 
\textit{Right:} IR band corresponding to the N-H bending mode of the NH$_{3}$ molecules (smoothed). The peak observed at $\sim$1500 cm$^{-1}$ 
corresponds to the formation of the NH$_2^{\cdot}$ radical.} 
\label{ir}
\end{figure*}

\begin{figure*}
\centering
\includegraphics[scale=0.45]{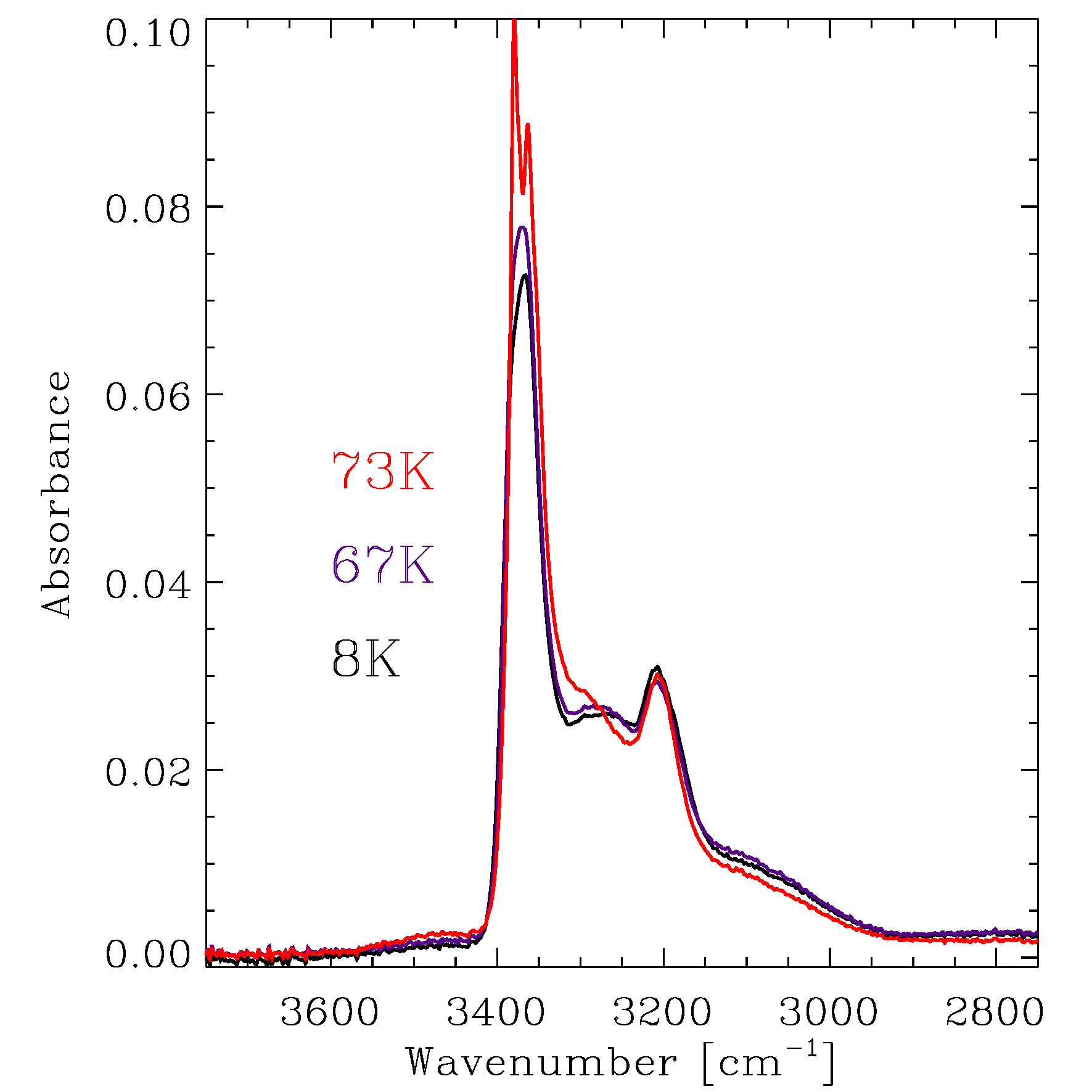}
\includegraphics[scale=0.45]{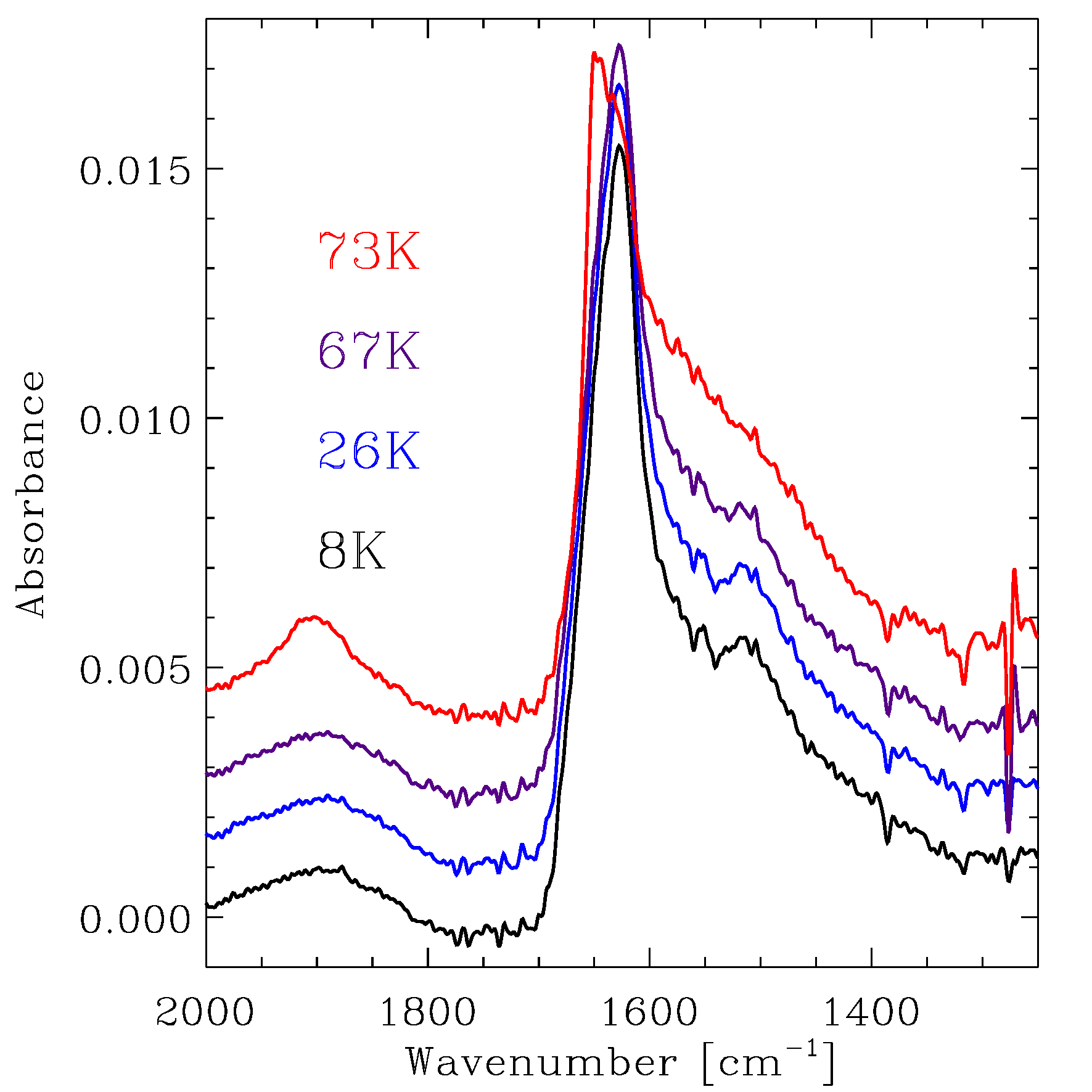}
\caption{IR spectra of the ice sample during the temperature-programmed desorption after irradiation 
in experiment 3 (see Table \ref{exps}). Similar results were found for the rest of experiments. 
\textit{Left:} Above 67 K, a change in the NH$_3$ ice matrix structure leads to a change in the N-H stretching band profile 
(see also Satorre et al. (2013), Giuliano et al. (2014)). 
At the same time, the shoulder above 3000 cm$^{-1}$ slightly decreases, probably due to the diffusion and subsequent reactions of 
the NH$^{\cdot}$ radical.
\textit{Right:} Above 26 K, the intensity of the peak corresponding to the NH$_2^{\cdot}$ radical also decreases due to its diffusion and 
subsequent reactions, and has completely disappeared at a temperature of 73 K, indicating that no free radicals are left in the ice 
sample at that temperature. Spectra are shifted for clarity.}
\label{ir-tpd}
\end{figure*}

\subsubsection{IR spectra of NH$_{3}$ ice during irradiation}
\label{results-ir}
The IR spectrum of a pure and amorphous NH$_3$ ice presents three different features: 
a wide band centered at $\sim$3300 cm$^{-1}$ (see black curve in left panel of Fig. \ref{ir}) corresponding to the N-H stretching 
modes (the symmetric mode peaks at $\sim$3375 cm$^{-1}$, and the antisymmetric mode at $\sim$3200 cm$^{-1}$); 
a weaker band at $\sim$1625 cm$^{-1}$ (see black curve in right panel of Fig. \ref{ir}) corresponding to the N-H bending mode; 
and a third band at 1070 cm$^{-1}$ (not shown) corresponding to the so-called umbrella mode. 
During photoprocessing of pure NH$_3$ ices, photodissociation and photodesorption of the NH$_3$ molecules decrease the absorbance of 
these bands, as shown in Fig. \ref{ir} for the former two features. 

Photodissociation in the gas phase of NH$_3$ molecules readily leads to the formation of NH$_2^{\cdot}$ and NH$^{\cdot\cdot}$ radicals due to the loss of one or two 
hydrogen atoms, respectively, although the formation of ground-state NH radicals is spin-forbidden (\cite{okabe}). 
Formation of NH$_2^{\cdot}$ during irradiation of pure NH$_3$ ice was reported in \cite{gerakines96} and \cite{loeffler10} thanks to an 
IR peak detected above 1500 cm$^{-1}$ on the red wing of the NH$_3$ bending band. This peak is observed in the right panel of 
Fig. \ref{ir}. 
In addition, a decrease in the intensity ratio between the symmetric and antisymmetric N-H stretching modes in the left panel of Fig. \ref{ir} 
is probably due to the contribution of the absorption corresponding to the antisymmetric N-H stretching mode of NH$_2^{\cdot}$ (\cite{milligan65}). 
At the same time, a wide shoulder at $\sim$3100 cm$^{-1}$  on the red wing of the N-H stretching band shown in the left panel of Fig. \ref{ir} 
could be attributed to the imidogen radical (NH$^{\cdot\cdot}$; \cite{rosengren65}). 
During the warm-up phase of the experiment, after photoprocessing of the ice sample is complete, the mobility of these radicals 
is increased. Diffusion and subsequent reaction of the radicals led to a decrease of their IR features. 
The IR peak at $\sim$1500 cm$^{-1}$ corresponding to the bending mode of the NH$_2^{\cdot}$ radical, disappears between 26 K and 73 K, as 
shown in the right panel of Fig. \ref{ir-tpd}. At 67 K, the intensity of this feature is roughly half of that at 8 K, 
and has completely disappeared at 73 K. 
In addition, a slight decrease in the shoulder at $\sim$3100 cm$^{-1}$ corresponding to the imidogen radical is observed at temperatures 
above 67 K (left panel of Fig. \ref{ir-tpd}). 

During irradiation, NH$^{\cdot\cdot}$ and NH$_2^{\cdot}$ radicals can be dissociated by subsequent photons leading to the production of H and N atoms, 
that recombine to form H$_2$ and N$_2$ molecules (see \cite{loeffler10}, where additional pathways to the formation of these photoproducts 
are reported), which are IR inactive due to the lack of an electric dipole moment. 
Formation of these molecules is confirmed thanks to the temperature-programmed desorption (TPD) of the irradiated ices 
during the warm-up phase of the experiments (see Sect. \ref{results-tpd}). 
In addition, formation of hydrazine (N$_2$H$_4$) by recombination of two NH$_2^{\cdot}$ radicals, or reaction of a NH$^{\cdot\cdot}$ radical 
with a NH$_3$ molecule (\cite{loeffler10}), is also reported at a temperature of 10 K in \cite{gerakines96}. 
Although no IR absorption corresponding to this molecule was detected during our experimental simulations, formation of this 
molecule was also confirmed during the TPD of the irradiated ices (Sect. \ref{results-tpd}). 

\begin{figure*}
\centering
\includegraphics[scale=0.4]{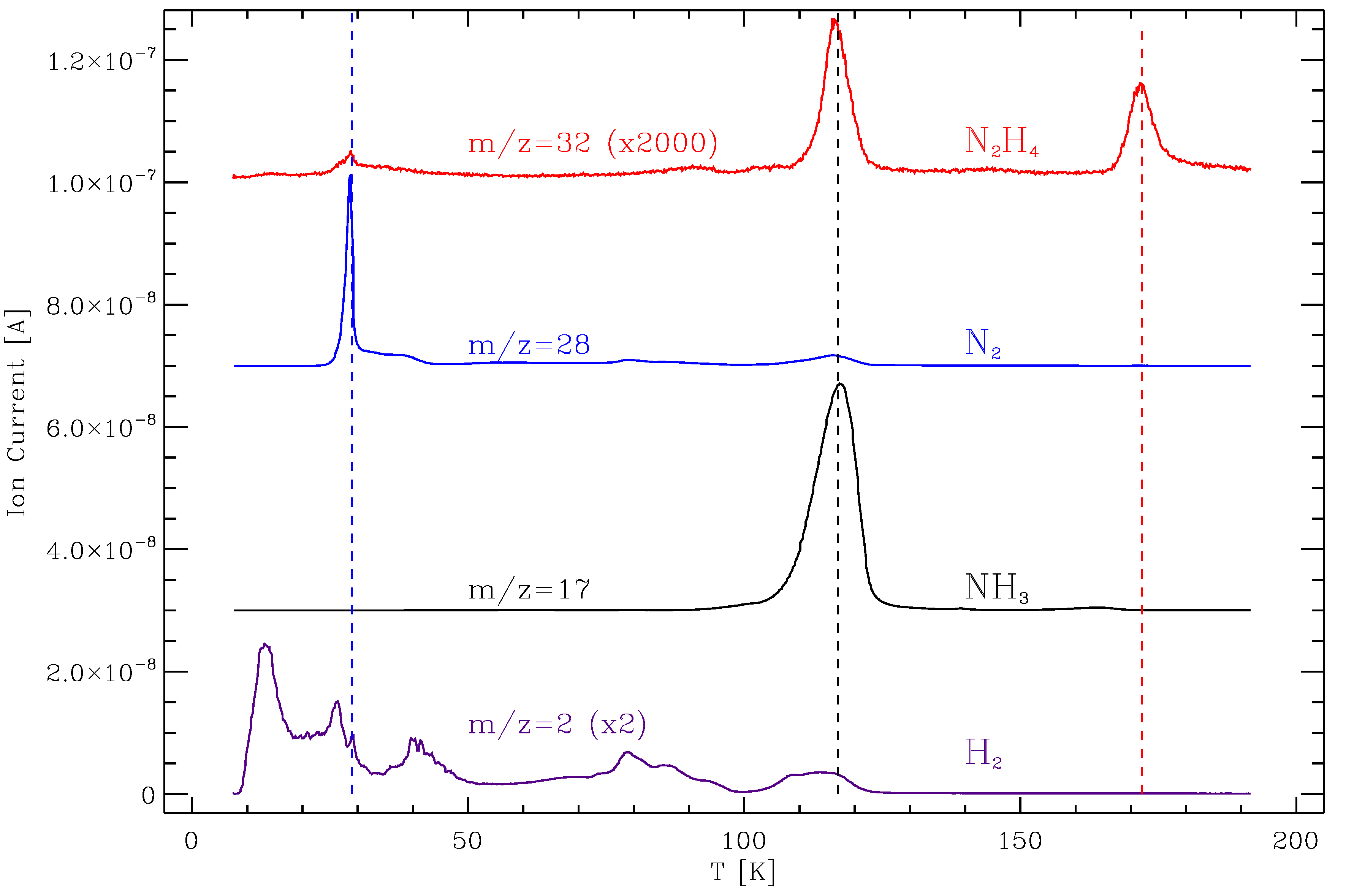}
\caption{Thermal desorption peaks detected by the QMS for different $m/z$ fragments during the TPD phase of experiment 6 
(see Table \ref{exps}, similar results were found for the rest of experiments) corresponding to 
NH$_{3}$ ($m/z$=17, black) and the three irradiation products: H$_2$ ($m/z$=2, purple), N$_2$ ($m/z$=28, blue), and N$_2$H$_4$ ($m/z$=32, red).
TPD curves are shifted for clarity.
The desorption peak observed for $m/z$=32 at T = 117 K is probably due to an ion-molecule reaction of ammonia inside the mass spectrometer, 
since it is also observed in experiments with no ammonia-ice irradiation.}
\label{tpd}
\end{figure*}

\subsubsection{Temperature-programmed desorption of the irradiated NH$_{3}$ ice}
\label{results-tpd}
After photoprocessing of the ice samples, a constant heating rate was applied to perform the TPD of the irradiated ices. 
Thermal desorption of ammonia and the photoproducts H$_2$, N$_2$, and N$_2$H$_4$ was detected by the QMS at a given temperature 
according to their volatility, as shown in Fig. \ref{tpd}. 
In particular, desorption of H$_2$ molecules was observed at a wide range of temperatures, since the very beginning of the warm-up 
phase of the experiment, due to the high volatility of this species. 
A desorption peak corresponding to most of the photoproduced N$_2$ molecules was observed at 29 K (slightly higher than the desorption 
temperature reported in \cite{fayolle16} for pure N$_2$ ices, probably due to the effect of the NH$_3$ matrix).  
On the other hand, 
the hydrazine 
molecules finally desorbed at 172 K, a desorption temperature 10-20 K lower than that reported in \cite{roux83} for experiments 
performed in a high-vacuum chamber with a base pressure five orders of magnitude higher compared to the one in the ISAC setup.

\begin{figure*}
\centering
\includegraphics[width=12cm]{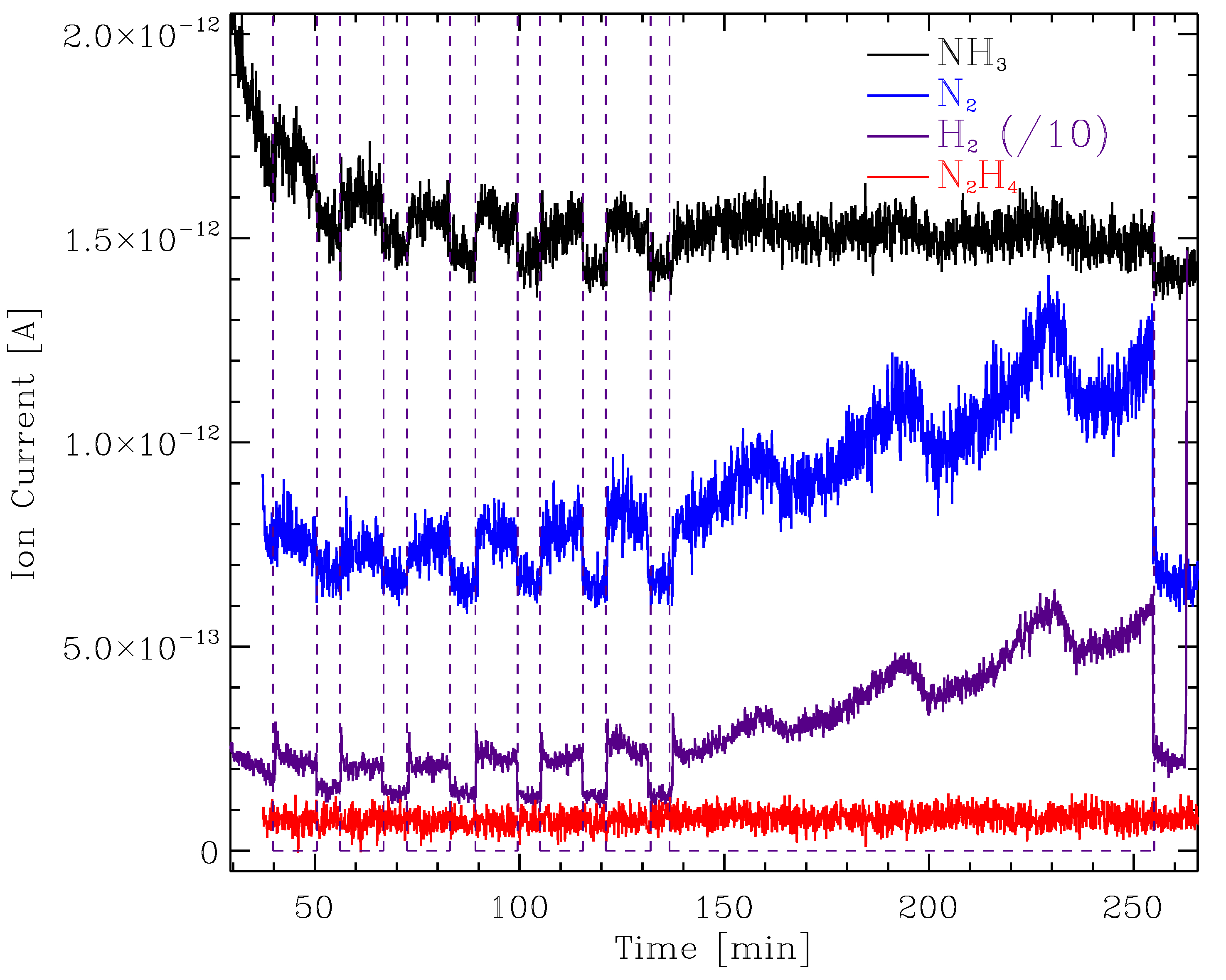}
\caption{Photodesorption of NH$_3$ (black), and the photoproducts H$_2$ (purple) and N$_2$ (blue) detected by the QMS during 
irradiation of the ice sample in experiment 1 (see Table \ref{exps}, similar results were found for the rest of experiments). 
Irradiation intervals are indicated with vertical dashed lines. 
The increase in the photodesorption of H$_2$ and N$_2$ suggests that these molecules photodesorb mainly through an indirect 
mechanism (see text), while the photoproduct N$_2$H$_4$ (red) was not found to photodesorb. 
Signals are shifted for clarity. 
Fluctuations in the signals are due to instabilities on the VUV flux during long irradiation periods.}
\label{pd}
\end{figure*}

\begin{table*}
\centering
\caption{Photodesorption yields measured for the different irradiation intervals during photoprocessing of a pure NH$_3$ ice in experiment 1, 
see Table \ref{exps}. 
Similar results were found  for experiments 2-6.}
\label{ypd}
\begin{tabular}{cccc}
\hline
\hline
Irradiation period&Fluence$^{a}$&$Y_{pd}$ (NH$_{3}$)$^{b}$&$Y_{pd}$ (N$_2$)$^{b}$\\
(min)&$\times$ 10$^{17}$ (photons cm$^{-2}$)&\multicolumn{2}{c}{$\times$ 10$^{-3}$ ($\frac{\text{molecules}}{\text{incident photon}}$)}\\
\hline
0 - 10&1.2&2.05&1.51\\
10 - 20&2.4&2.12&1.36\\
20 - 30&3.6&1.89&2.18\\
30 - 40&4.8&2.28&2.41\\
40 - 50&6.0&1.92&2.64\\
50 - 60&7.2&2.01&3.05\\ 
60 - 180&21.6&1.74&5.86\\
\hline
\end{tabular}
\begin{list}{}
\item $^{a}$ Total fluence at the end of every irradiation period. 
\item $^{b}$ Averaged for every irradiation period. 
A factor of 2 is assumed as the error in the photodesorption yield values due to the uncertainties in all the parameters 
of equation \ref{eqmscal} (see also \cite{martin16}).
\end{list}
\end{table*}

\subsection{Photodesorption from pure NH$_{3}$ ice}
\label{results-pd}
During irradiation of the ice samples, the photodesorbing molecules were detected in the gas phase by the QMS. 
In particular, photodesorption led to an increase on the signal of the m/z fragments of the photodesorbing 
molecules during every irradiation interval. 
The measured ion current of the main mass fragments corresponding to NH$_3$ (m/z = 17), and the three photoproducts: H$_2$ (m/z = 2), 
N$_2$ (m/z = 28), and N$_2$H$_4$ (m/z = 32), are shown in Fig. \ref{pd}. 
Photodesorption of the former three species was detected, as in \cite{nishi84} and \cite{loeffler10}, 
while N$_2$H$_4$ was not found to photodesorb. 
The integrated QMS ion currents were converted into photodesorbing column densities using Equations \ref{eqmscal} and \ref{eqmscorr}, 
the parameter values shown in Table \ref{param}, and a $k_{CO}$ = 5.55 $\times$ 10$^{-12}$ A min ML$^{-1}$. 
The ratio of the photodesorbing column densities in every irradiation period and the incident fluence during the same period 
led to the average photodesorption yield for every irradiation interval shown in Table \ref{ypd}. 
Quantification of the photodesorbing H$_2$ molecules was not included since the measured values suffered from background H$_2$ 
contamination, which is the most usual contaminant in UHV chambers. 

\smallskip

Photodesorption of NH$_3$ molecules took place with an average photodesorption yield of 2.1 x 10$^{-3}$ $\frac{\text{molecules}}{\text{incident photon}}$, 
that remained constant during the whole experiment (see Table \ref{ypd}). 
This is similar, within errors, to the photodesorption yield indirectly measured in the solid phase by \cite{loeffler10} 
with a quartz-crystal microbalance (QCM)\footnote{The QCM measured the total ice mass loss, which was assigned mostly to ejected NH$_3$ 
molecules, although it included any other photodesorbing product.} during photoprocessing of a pure NH$_3$ ice grown at 40 K using a 193 nm ArF laser. 
%

During irradiation of pure ice samples, it is not possible to elucidate whether photodesorption of the original ice component 
takes place through a direct mechanism or 
through a mechanism involving energy transfer attending only to the evolution of the photodesorption yield with fluence. 
This can only be done 
for the photodesorption of photoproducts, as explained in Sect. \ref{intro}. 
\cite{nishi84} measured the energy distribution of the photodesorbing NH$_3$ molecules from a pure NH$_3$ ice grown at 90 K and 
irradiated with a 193 nm ArF laser, 
using a time-of-flight (TOF) mass spectrometer. 
Two energy components with mean translational energies of 0.17 eV and 0.65 eV, respectively, were found. 
The low energy component was associated to NH$_3$ molecules photodesorbing through a mechanism involving energy transfer 
from the absorbing molecule in the bulk of the ice to a surface molecule. This is similar to the DIET mechanism involving energy 
transfer described in Sect. \ref{intro}. 
In this case, absorption of a photon led to a transition to a Wannier exciton state that subsequently propagated through the crystal 
approaching a weakly bound edge molecule that acquired linear momentum due to the electronic repulsive force. 
On the other hand, the high energy component, that accounted for half of the photodesorbing molecules, was associated to excited 
surface molecules, probably desorbing after recombination of a previously photodissociated NH$_3$ molecule. 

\smallskip

Photodesorption of H$_2$ and N$_2$ molecules took place with an increasing photodesorption yield with fluence, as reported in 
\cite{loeffler10}. 
The photodesorption yield of N$_2$ was found to be  1.4 x 10$^{-3}$ $\frac{\text{molecules}}{\text{incident photon}}$ 
(i.e., on the order of that of NH$_3$) 
for a fluence of up to 2.4 $\times$ 10$^{17}$ photons cm$^{-2}$, and then gradually raised,  
being $\sim$4 times higher when the total fluence was 2.2 $\times$ 10$^{18}$ (see Table \ref{ypd}).  
In this case, the increasing photodesorption yield is related to the accumulation of photoproduced H$_2$ and N$_2$ molecules in the 
ice sample prior to their desorption into the gas phase. 
Therefore, there is a significant contribution of indirect photodesorption mechanisms involving energy 
transfer. These mechanisms require the absorption of a new photon after the formation of the photoproducts, as explained in Sect. \ref{intro}, 
in contrast to the direct photodesorption mechanisms where desorption occurs immediately after the formation of the molecules, 
without the need of absorbing a new photon. In that case, the photoproduct molecules could not accumulate on the surface of the ice 
before their desorption, leading to a constant photodesorption yield with fluence 
(see \cite{martin15,martin16,gustavo16,gustavo17}). 

\begin{table*}
\centering
\caption{Photodesorption yields measured for pure ices during irradiation with a MDHL.}
\label{comp}
\begin{tabular}{ccc}
\hline
\hline
Ice component&$Y_{pd}$ & Reference\\
&$\times$ 10$^{-3}$ ($\frac{\text{molecules}}{\text{incident photon}}$)&\\
\hline
H$_2$O&1.3 $\pm$ 0.2&\cite{gustavo17}\\
CO&54.0 $\pm$ 5.0&\cite{guillermo10}\\
CO$_2$&0.1 $\pm$ 0.04&\cite{martin15}\\
CH$_3$OH&$\le$0.03&\cite{gustavo16}\\
NH$_3$&2.0$^{+2.1}_{-1.0}$&This work\\
\hline
\end{tabular}
\end{table*}

\section{Astrophysical implications}
\label{imp}

Ammonia ice is thought to form early at the onset of the diffuse cloud collapse that leads to the formation of dense clouds where 
ice mantles are usually detected, by hydrogenation reactions of N atoms on the surface of dust grains. 
The abundance of NH$_3$ with respect to H$_2$O in the ice mantles is around 5\% in dense clouds and also in the cold circumstellar 
envelopes around protostars (see \cite{boogert15}, and references therein). 
Its importance, though, is beyond all doubts since it is the only confirmed source of N atoms in interstellar ices. 
The detection of ammonia in the gas-phase of the cold ISM was already reported in \cite{cheung68}. 

NH$_3$ ice molecules are photoprocessed along with all the ice components in the interior of dense clouds thanks to the secondary 
UV field mentioned in Sect. \ref{intro}, leading to photochemical reactions that produce new ice molecules 
and photodesorption processes that allow the presence of molecules in the gas phase of cold interstellar regions where 
thermal desorption is negligible. 
These processes can be simulated in the laboratory under astrophysically relevant conditions. 
For the VUV flux of the lamp used (see Sect. \ref{methods}),  
the ice samples experience a fluence of $\sim$3 $\times$ 10$^{17}$ photons cm$^{-2}$ after $\sim$30 minutes 
of irradiation in our experiments, which is similar to the fluence experienced by the ice mantles during the 
expected lifetime of a molecular cloud, assuming a secondary UV flux of $\sim$10$^4$ photons cm$^{-2}$ s$^{-1}$(\cite{shen04}).  
The six components confirmed to be present in the interstellar ices are distributed over two distinct ice layers as the result 
of the formation process (see \cite{boogert15}, and references therein). 
A polar ice layer is formed first on the surface of dust grains, dominated by H$_2$O and including NH$_3$, CH$_4$, and 
a fraction of the CO$_2$ ice molecules. 
On top of this polar ice layer, an apolar layer is subsequently formed, mainly composed by CO along with the rest 
of the CO$_2$ molecules and probably CH$_3$OH. 
In any case, irradiation experiments of pure ices are used as benchmarks in the study of the photoprocessing of more realistic 
multicomponent ices. 

In our experimental simulations using a MDHL, whose emission spectrum is similar to that expected in the interior of dense clouds, 
photodesorption of NH$_3$ molecules is observed to proceed with 
a constant yield with fluence during irradiation of a pure NH$_3$ ice  
($Y_{pd}$ (NH$_3$) = 2.1$^{+2.1}_{-1.0}$ x 10$^{-3}$ $\frac{\text{molecules}}{\text{incident photon}}$). 
This value is on the order of the observed H$_2$O photodesorption during photoprocessing of pure H$_2$O ices using the same UV lamp, 
(\cite{gustavo17}), while it is one order of magnitude lower than the photodesorption yield 
of CO molecules during photoprocessing of a pure CO ice (\cite{guillermo10}), and one order of magnitude higher than that measured 
for CO$_2$ (see Table \ref{comp}) under similar conditions (\cite{martin15}). 
Photodesorption of NH$_3$ molecules from an ice mixture dominated by water will be addressed in a forthcoming paper, and 
is expected to be lower than the value presented in this paper for segregated (pure) NH$_3$ ice (\cite{loeffler10}). 

On the other hand, we also observed photodesorption of the produced N$_2$ molecules, with a yield that increased for fluences higher 
than 2.4 $\times$ 10$^{17}$ photons cm$^{-2}$. 
For a fluence similar to that experienced by the ice mantles in the expected lifetime of a dense cloud 
($\sim$3 $\times$ 10$^{17}$ photons cm$^{-2}$), the average photodesorption 
yield of N$_2$ molecules was similar to that observed for the NH$_3$ molecules 
($Y_{pd}$ (N$_2$) = 1.7$^{+1.7}_{-0.9}$ x 10$^{-3}$ $\frac{\text{molecules}}{\text{incident photon}}$).  
Photodesorption of N$_2$ molecules from a pure N$_2$ ice is expected to be negligible, due to the low VUV-absorption cross section 
of this ice (the average absorption cross section is 7.0 $\times$ 10$^{-21}$ cm$^{2}$, 
while for NH$_3$ ices it is 6.1 $\times$ 10$^{-18}$ cm$^{2}$; \cite{gustavo14a,gustavo14b}). 
Detection of gas-phase N$_2$ was reported in \cite{knauth04} from far-UV observations toward the star HD 124314, 
with an abundance that could not be explained with gas-phase chemical models for either diffuse or dense clouds. 

\section{Conclusions}
\label{conclusions}
We have performed experimental simulations of the UV photoprocessing of NH$_3$ ice using a MDHL with an emission spectrum similar 
to that expected to be present in the interior of dense clouds. 
IR spectroscopy in transmittance was used to monitor the solid sample during irradiation. 
Formation of NH$_2^{\cdot}$ and NH$^{\cdot\cdot}$ radicals due to the photodissociation of NH$_3$ molecules was observed thanks to the appareance of 
IR features at $\sim$1500 cm$^{-1}$ and $\sim$3100 cm$^{-1}$, respectively. 
The photoproducts H$_2$ and N$_2$ are not IR active, and the N$_2$H$_4$ IR features did not reach the sensitivity limit in 
the spectra of the irradiated ices, but 
they were detected in the gas phase after thermal desorption during the TPD of the photoprocessed ice thanks to a QMS. 

In addition, photodesorption of NH$_3$ and the photoproducts H$_2$ and N$_2$ were detected by the QMS during photoprocessing 
of the ice samples. 
Photodesorption of NH$_3$ took place with a constant yield with fluence, 
as it is the case for other species during irradiation of their pure ices. 
On the other hand, the photodesorption yield of H$_2$ and N$_2$ were observed to increase with fluence, 
pointing toward a significant contribution of indirect photodesorption mechanism involving energy transfer from the absorbing 
molecule to a previously formed H$_2$ or N$_2$ molecule, that subsequently desorbed. 
This kind of mechanism allows accumulation of the photoproducts prior to their desorption, leading to an increase of the 
number of molecules available for the photodesorption and, therefore, an increase of the photodesorption yield with fluence. 

The calibration of the QMS allowed us to quantify the photodesorption yields for the desorbing molecules. 
NH$_3$ molecules were found to photodesorb with an average yield of 2.1$^{+2.1}_{-1.0}$ x 10$^{-3}$ $\frac{\text{molecules}}{\text{incident photon}}$, 
which is of the same order than the photodesorption measured for H$_2$O molecules during irradiation of a pure water ice (\cite{gustavo17}). 
N$_2$ molecules are not expected to efficiently photodesorb from pure N$_2$ ices 
($Y_{pd}$ (N$_2$) $\le$ 2 $\times$ 10$^{-4}$ $\frac{\text{molecules}}{\text{incident photon}}$; \cite{oberg09b}), 
probably due to their low VUV-absorption cross section (\cite{gustavo14b}). 
During irradiation of a pure NH$_3$ ice, a photodesorption yield similar to that of NH$_3$ molecules was measured for N$_2$ 
molecules, for a fluence equivalent to that experienced by ice mantles 
during the expected lifetime of dense clouds ($Y_{pd}$ (N$_2$) = 1.7$^{+1.7}_{-0.9}$ x 10$^{-3}$ $\frac{\text{molecules}}{\text{incident photon}}$), 
although photodesorption yields up to four times higher were measured for higher fluences.

\section*{Acknowledgments}
This research was financed by the Spanish MINECO under projects AYA2011-29375 and AYA2014-60585.

\bsp	
\label{lastpage}
\end{document}